\documentclass[twoside]{article}

\usepackage{color}
\usepackage{amssymb}
\usepackage{dsfont}
\usepackage{amssymb,amsmath,ulem,cancel}

\numberwithin{equation}{section}

\newtheorem{theorem}{Theorem}[section]
\newtheorem{lem}{Lemma}[section]
\newtheorem{pro}{Proposition}[section]
\newtheorem{cor}{Corollary}[section]
\newtheorem{rem}{Remark}[section]
\newtheorem{rems}{Remarks}[section]
\newtheorem{ex}{Example}[section]
\newtheorem{defi}{Definition}[section]
\newtheorem{hyp}{Assumption}[section]
\newtheorem{con}{Conjecture}[section]

\newcommand{\bt}{\begin{theorem}}
\newcommand{\et}{\end{theorem}}
\newcommand{\bl}{\begin{lem}}
\newcommand{\el}{\end{lem}}
\newcommand{\bp}{\begin{pro}}
\newcommand{\ep}{\end{pro}}
\newcommand{\bcor}{\begin{cor}}
\newcommand{\ecor}{\end{cor}}
\newcommand{\bcon }{\begin{con} \rm }
\newcommand{\econ }{\end{con}}
\newcommand{\lab }{\label }
\newcommand{\bd}{\begin{defi} \rm }
\newcommand{\ed}{\end{defi}}
\newcommand{\brem }{\begin{rem} \rm }
\newcommand{\erem }{\end{rem}}
\newcommand{\brems }{\begin{rems} \rm }
\newcommand{\erems }{\end{rems}}
\newcommand{\bhyp }{\begin{hyp} \rm }
\newcommand{\ehyp }{\end{hyp}}
\newcommand{\bex}{\begin{ex} \rm }
\newcommand{\eex}{\end{ex}}
\newcommand{\be}{\begin{equation}}
\newcommand{\ee}{\end{equation}}
\newcommand{\bde}{\begin{displaymath}}
\newcommand{\ede}{\end{displaymath}}
\newcommand{\beq}{\begin{eqnarray*}}
\newcommand{\eeq}{\end{eqnarray*}}
\newcommand{\beqa}{\begin{eqnarray}}
\newcommand{\eeqa}{\end{eqnarray}}
\newcommand{\bea}{\begin{align*}}
\newcommand{\eea}{\end{align*}}

\def\proof{\noindent {\it Proof. $\, $}}
\def\endproof{\hfill $\Box$ \vskip 5 pt}

\def\I{\mathds{1}}
\def\wh{\widehat}
\def\wt{\widetilde}

\def\phi{\varphi }

\newcommand{\Sst}{S^{s,t}}
\newcommand{\aass}{\mbox{\rm a.s.}}
\newcommand{\aaee}{\mbox{\rm a.e.}}

\newcommand{\ssx}{s}
\newcommand{\ssy}{s}

\newcommand{\Leb }{\ell }

\newcommand{\pA}{A}
\newcommand{\pC}{C}

\newcommand{\pCc}{c}

\newcommand{\etab}{\eta^b}
\newcommand{\etal}{\eta^l}

\newcommand{\Blr}{B^{l}}
\newcommand{\Bbr}{B^{b}}

\newcommand{\rll}{r^{l}}
\newcommand{\rbb}{r^{b}}

\newcommand{\Vnet}{{V}^{\textrm{net}}}

\def\t1{\tau_{(1)}}

\def\gg{{\mathbb G}}

\def\G{{\cal G}}

\def\VLL{V^0}
\def\P{\mathbb P}

\def\PT{\wt {\mathbb P}}
\def\PTb{\wt {\mathbb P}^{\beta }}

\def\EP{{\mathbb E}_{\mathbb P}}

\def\C-FVA{{\rm C-FVA}}

\newcommand{\VCc}{V^C}

\newcommand{\wHzero}{\wh{\mathcal{H}}_{0}^{2}}
\newcommand{\wHzerd}{\wh{\mathcal{H}}_{0}^{2,d}}

\newcommand{\sumik}{\textstyle{\sum}}

\newcommand{\Keywords}[1]{\par\noindent{\small{\bf Keywords\/}: #1}}
\newcommand{\Class}[1]{\par\noindent{\small{\bf Mathematics Subjects Classification (2010)\/}: #1}}

\title{{\Large \bf FAIR BILATERAL PRICES IN BERGMAN'S MODEL} \vskip 105 pt }

\author{Tianyang Nie and Marek Rutkowski\footnote{The research of Tianyang Nie and Marek Rutkowski
was supported under Australian Research Council's Discovery Projects funding scheme (DP120100895).}
\\ School of Mathematics and Statistics \\ University of Sydney
\\ Sydney, NSW 2006, Australia}

\date{\vskip 70 pt  1 December 2014 \vskip 70 pt}

\begin{document}

\maketitle

\begin{abstract}
In this paper, we examine the pricing and hedging of a contract in the model proposed by Bergman \cite{B-1995} from the perspective of the hedger and his counterparty with arbitrary initial endowments. We derive inequalities satisfied by unilateral prices of a contract and we give the range for its fair bilateral prices. Our study hinges on results for BSDE driven by a multi-dimensional continuous martingales obtained in \cite{NR3}. We also derive the pricing PDEs for path-independent contingent claims of European style in a Markovian framework.
\vskip 20 pt
\vskip 20 pt
\Class{91G20,$\,$91G80}
\end{abstract}


\newpage

\section{Introduction} \label{sect1}

In Bielecki and Rutkowski \cite{BR-2014}, the authors introduced a generic nonlinear market model which includes several risky assets, multiple funding accounts and margin accounts (for related studies by other authors, see also \cite{BCPP11,BK09,BK11,SC12a,SC12b,PPB12,P10}). Using a suitable version of the no-arbitrage argument, they first discussed the hedger's fair price for a contract in the market model without collateralization (see Section 3.2 in \cite{BR-2014}). Subsequently, for a collateralized contract that can be replicated, they defined the hedger's ex-dividend price (see Section 5 in \cite{BR-2014}). It was also shown in \cite{BR-2014} that the theory of backward stochastic differential equations (BSDEs) is an important tool to compute the ex-dividend price (see, e.g., Propositions 5.2 and 5.4 in \cite{BR-2014}). It is worth mentioning that all the pricing and hedging arguments in \cite{BR-2014} are given from the viewpoint of the hedger and no attempt was made there to derive no-arbitrage bounds for unilateral prices.

We consider the problem of pricing and hedging of a derivative contract from the perspective of the hedger and his counterparty. Since we work within a nonlinear trading set-up, where the nonlinearity stems from the different interest rates and collateralization, the hedger's and counterparty's price do not necessarily coincide. Therefore, our goal is to compare the hedger's and counterparty's prices and to derive the range for no-arbitrage prices.  As shown by Bergman \cite{B-1995}, in the model with different lending and borrowing rates, which is a relatively simple instance of a nonlinear market model, the no-arbitrage price of any contingent claims must belong to an arbitrage band with the upper (resp., lower) bound given by the hedger's  (resp., the counterparty's) price of the contract.  In a recent paper by Mercurio \cite{M-2013}, the author extended some results from \cite{B-1995} by examining the pricing of European options in a model with different lending and borrowing interest rates and under collateralization.  As emphasized in related papers \cite{BR-2014,NR2,NR3}, in the nonlinear setup, especially in the market with different interest rates and idiosyncratic funding costs for risky assets, the initial endowments of the hedger and the counterparty are important. Unlike in the classic options pricing model, which enjoys linearity, it is no longer sufficient to consider the case of null initial endowments since the ex-dividend prices may depend on initial endowments (see Proposition 5.2 in \cite{BR-2014}). Therefore, the results obtained in \cite{B-1995} and \cite{M-2013} are only valid in  situation where the initial endowments of the hedger and the counterparty are assumed to be null.

We revisit the market model studied by Bergman \cite{B-1995} and we extend it in several respects.
First, we study general collateralized contracts, rather than path-independent European claims.
Second, we assume that investors have possibly non-zero (either positive or negative) initial endowments and both parties are allowed to use their initial endowments to invest in risky assets for the purpose of hedging.
Finally, we do not assume a priori any particular financial model, but rather we work within an abstract semimartingale
set-up. Our main goals are to examine how the initial endowment affects the price and to establish the existence of a non-empty interval for fair bilateral prices. We argue that the properties of their respective prices will be quite different
under alternative assumptions about initial endowments of both parties. As in \cite{NR2}, we show that the pricing inequalities can be obtained from the general results for the non-linear BSDEs, which determine unilateral prices and hedging strategies for both parties. For the sake of completeness, we also derive the pricing PDEs for path-independent European claims in a Markovian framework, thus extending once again the approach of Bergman \cite{B-1995}.

This work is organized as follows. In Section  \ref{sect2}, we introduce our set-up and we recall definitions and results regarding hedging strategies for collateralized
contracts in a model with different lending and borrowing rates. We also show there that the model is arbitrage-free for both
parties, in the sense of Definition \ref{arbitrage using nettled wealth}. For a more extensive discussion of models with funding costs and collateralization, the reader is referred to \cite{BR-2014,NR2}. In Section  \ref{sect3},  we first establish the existence and uniqueness of solutions to BSDEs yielding the ex-dividend prices and hedging strategies for the hedger and the counterparty. Next, we apply the comparison theorem for BSDEs driven by a multi-dimensional martingale established in \cite{NR3} to derive the range for fair bilateral prices.
In Section  \ref{sect4}, we place ourselves in a Markovian framework and we postulate that the interest rates are deterministic. Using the non-linear version of the Feynman-Kac formula, we derive the pricing PDEs for both parties and we describe their respective hedging strategies in terms of solutions to these PDEs.

\newpage

\section{Trading under Differential Rates and Collateralization}   \label{sect2}

Throughout the paper, we fix a finite trading horizon date $T>0$ for our model of the financial market.
Let $(\Omega, \G, \gg , \P)$ be a filtered probability space satisfying the usual conditions of right-continuity and completeness, where the filtration $\gg = (\G_t)_{t \in [0,T]}$ models the flow of information available to all traders. For convenience, we assume that the initial $\sigma$-field ${\cal G}_0$ is trivial. Moreover, all processes introduced in what follows are implicitly assumed to be $\gg$-adapted  and any semimartingale is assumed to be c\`adl\`ag.

\noindent {\bf Risky assets.} For $i=1,2, \dots, d$, we denote by $S^i$ the {\it ex-dividend price} of the $i$th risky asset with the {\it cumulative dividend stream} $\pA^i$. The process $S^i$ is aimed to represent the price of any traded security, such as, stock, stock option, interest rates swap, currency option, cross-currency swap, CDS, CDO, etc.

\noindent {\bf Cash accounts.} The riskless  {\it lending} (resp., {\it borrowing}) {\it cash account} $B^l$ (resp., $B^b$) is used for unsecured lending (resp., borrowing) of cash.

\bhyp \lab{assumption for primary assets}
The price processes of {\it primary assets} are assumed to satisfy: \hfill \break
(i) For each $i=1,2,\dots , d$, the price $S^i$ is semimartingale and the cumulative dividend stream $\pA^i$ is finite variation process with $\pA^i_{0}=0$.\hfill \break
(ii) The riskless accounts $B^{l}$ and $B^{b}$ are strictly positive and continuous processes
of  finite variation with $B^l_0=B^b_0 =1$ for $i=1,2\dots , d$.
\ehyp

 By a {\it bilateral financial contract}, or simply a {\it contract}, we mean an arbitrary c\`adl\`ag process $\pA$ of finite variation. The process $A$ is aimed to represent the {\it cumulative cash flows} of a given contract from time 0 till its maturity date $T$. By convention, we set $\pA_{0-}=0$.

The process $\pA$ is assumed to model all cash flows of a given contract, which are either paid out from the wealth or added to the wealth, as seen from the perspective of the {\it hedger} (recall that the other party is referred to as the {\it counterparty}).
Note that the process $A$ includes the initial cash flow $A_0$ of a contract at its inception date $t_0=0$.
For instance, if a contract has the initial {\it price} $p$ and stipulates that the hedger will receive cash flows  $\bar{\pA}_1,
\bar{\pA}_2, \dots , \bar{\pA}_k$ at times $t_1, t_2, \dots , t_k \in (0,T]$, then we set $A_0=p$ so that
\bde
\pA_t = p + \sum_{l=1}^k \I_{[t_l,T]}(t) \bar{\pA}_l .
\ede
The symbol $p$ is frequently used to emphasize that all future cash flows $\bar{\pA}_l$ for $l=1,2, \dots, k$ are explicitly specified by the contract's covenants, but the initial cash flow $A_0$ is yet to be formally defined and evaluated.
Valuation of a contract $A$ means, in particular, searching for the range of {\it fair values} $p$ at time $0$ from the viewpoint of either the hedger or the counterparty. Although the valuation paradigm will be the same for the two parties, due either to the asymmetry in their trading costs and opportunities, or the non-linearity of the wealth dynamics, they will typically obtain different sets of fair prices for~$A$. This is the main objective of our current work.

\subsection{Collateralization} \label{sect2.1}

In this paper, we examine the situation when the hedger and the counterparty enter a contract and either receive or post collateral with the value formally represented by an exogenously given stochastic process~$\pC $, which is assumed to be a semimartingale (or, at least, a c\`adl\`ag process). The process $C$ is referred to as either the {\it margin account} or the {\it collateral amount}. Let
\be \lab{collss}
\pC_t = \pC_t \I_{\{ \pC_t \geq 0\}} +  \pC_t \I_{\{ \pC_t < 0\}} = \pC^+_t - \pC^-_t.
\ee
By convention, $\pC^+_t$ is the cash value of collateral received at time $t$ by the hedger, whereas $\pC^-_t$ represents the cash value of collateral posted by him. For simplicity of presentation, it is postulated throughout that only cash collateral may be posted or received (for other conventions, see \cite{BR-2014}).

We also make the following natural assumption regarding the state of the margin account at the contract's maturity date.

\bhyp \lab{assumption for margin account on maturity date}
The $\gg$-adapted collateral amount process $C$ satisfies $C_T=0$.
\ehyp

The equality $C_T=0$ ensures that any collateral amount posted is returned in full to its owner at the contract's expiration, provided that the default event does not occur at $T$. Of course, if the default event is also modeled, which is not the case in this work, then one needs to specify the closeout payoff as well.

\brem The current financial practice typically requires the collateral amounts to be held in {\it segregated} margin accounts,
so that the hedger, when he is a collateral taker, cannot make use of the collateral amount for trading.
Another collateral convention encountered  in practice is {\it rehypothecation}, which refers to the situation where a bank is allowed to reuse the collateral pledged by its counterparties as collateral for its own borrowing.
Note that if the hedger is a collateral giver, then a particular convention regarding segregation or rehypothecation is immaterial for the wealth dynamics of his portfolio.
\erem

We are in a position to introduce trading strategies based on a finite family of primary assets.
For simplicity, all issues are discussed from from the perspective of the hedger, unless explicitly stated otherwise.
It is clear that to cover the counterparty it suffices to replace $(A,C)$ by $(-A,-C)$. The following definition
is a special case of Definition 4.1 in \cite{BR-2014}

\bd \lab{tsx2}
A {\it collateralized hedger's trading strategy} is a quadruplet $(x,\phi , \pA , \pC )$ where a portfolio $\phi $, given by
\be \lab{vty}
\phi = \big( \xi^1,\dots , \xi^{d},\psi^{l},\psi^{b}, \etab, \etal,\psi^{d+1} , \eta^{d+2} \big)
\ee
is composed of the {\it risky assets} $S^i,\, i=1,2,\ldots,d$, the {\it unsecured lending cash account} $B^l$ the {\it unsecured borrowing cash account} $B^b$,  the {\it collateral accounts} $B^{\pCc,b}$ and $B^{\pCc,l}$, the {\it borrowing account} $B^{d+1}$  associated with the posted cash collateral, and the {\it lending account} $B^{d+2}$ associated with the received cash collateral.
\ed

For a detailed explanation of all terms arising in the definition of a strategy $\phi $, the reader is referred to Section 4.1 in \cite{BR-2014}. Let us only mention that if $B^{\pCc,b}\neq B^{\pCc,l}$,  for example if the hedger post the collateral, he will receives interest from the counterparty determined by $B^{\pCc,l}$, that is, the counterparty pays the hedger the interest determined by $B^{\pCc,l}$ not $B^{\pCc,b}$. This creates asymmetric financial environments for the hedger and the counterparty.
We make the following standing assumption.

\bhyp \lab{assumption for collateral account}
The accounts $B^{c,l},\, B^{c,b},\, B^{d+1}$ and $B^{d+2}$  are strictly positive, continuous processes
of finite variation with $B^{c,l}_0=B^{c,b}_0=B^{d+1}_0=B^{d+2}_0=1$.
\ehyp

The case of the {\it cash collateral} is described by the following postulates: \hfill \break
(i) If the hedger receives at time $t$ the amount $\pC^+_t$ as cash collateral, then he pays
to the counterparty interest determined by the amount $\pC^+_t$ and the account $B^{\pCc,b}$.
Under segregation, he receives interest determined by the amount $\pC^+_t$ and the account $B^{d+2}$ and thus $\eta^{d+2}_t B^{d+2}_t = C^+_t$.  When rehypothecation is considered,  the hedger may temporarily (i.e., before the contract's maturity date or the default time, whichever comes first) utilize the cash amount $\pC^+_t$ for trading and thus $\eta^{d+2}= 0$.
\hfill \break (ii) If the hedger posts a cash collateral at time $t$, then the collateral amount is borrowed from
the dedicated  collateral borrowing account $B^{d+1}$. He receives interest determined by the amount $\pC^-_t$ and the collateral account $B^{\pCc,l}$. We postulate that
\be \lab{565656}
\psi^{d+1}_t B^{d+1}_t = - C^-_t .
\ee

\subsection{Trading Strategies and Wealth Processes}    \label{sect2.2}

We examine trading from the perspective of the hedger with an arbitrary initial endowment. For the counterparty, we may use similar arguments after replacing $(A,C)$ by $(-A,-C)$.

In the context of a collateralized contract, we find it convenient to introduce:
\hfill \break (i) the process $V_t(x,\phi , \pA ,\pC)$ representing the hedger's wealth at time $t$,
\hfill \break (ii) the process $V_t^p(x,\phi , \pA ,\pC)$ representing the value of hedger's portfolio at time $t$,
\hfill \break (iii) the {\it adjustment process} $\VCc_t(x,\phi , \pA ,\pC) := V_t(x,\phi , \pA ,\pC) - V^p_t(x,\phi , \pA ,\pC)$, which is aimed to quantify the impact of the margin account on a trading strategy.

\bd \lab{ts2x}
The hedger's {\it portfolio's value} $V^p(x,\phi , \pA ,\pC)$ is given by
\be \lab{poutf1}
V^p_t (x,\phi , \pA ,\pC) =  \sum_{i=1}^{d} \xi^i_t S^i_t + \psi^l_t B^l_t + \psi^b_t B^b_t +  \psi^{d+1}_t B^{d+1}_t .
\ee
The hedger's {\it wealth} $V(x,\phi , \pA ,\pC)$ equals
\be \lab{portf1}
V_t (x,\phi , \pA ,\pC) =  \sum_{i=1}^{d} \xi^i_t S^i_t + \psi^l_t B^l_t+\psi^b_t B^b_t
+  \etab_t B^{\pCc,b}_t+ \etal_t B^{\pCc,l}_t +  \psi^{d+1}_t B^{d+1}_t +  \eta^{d+2}_t B^{d+2}_t  .
\ee
\ed

In general, the adjustment process $\VCc(x,\phi , \pA ,\pC)$ equals
\be \lab{portf1b}
\VCc_t (x,\phi , \pA ,\pC) = \etab_t B^{\pCc,b}_t+ \etal_t B^{\pCc,l}_t +  \eta^{d+2}_t B^{d+2}_t = - C_t +  \eta^{d+2}_t B^{d+2}_t
\ee
where $\etab_t =-  (B^{\pCc,b}_t)^{-1}\pC_t^+$ and $\etal_t =(B^{\pCc,l}_t)^{-1} \pC_t^-$.
In what follows, we only consider the case of cash collateral under rehypothecation, that is, we set $\eta^{d+2}= 0$.
Moreover, for simplicity of presentation, we assume that the collateral borrowing account $B^{d+1}$ coincides with $B^{b}$,
so that we may and do set $\psi^{d+1}=0$.

The self-financing property of the hedger's strategy is defined in terms of the dynamics of the value process.
Note that we use here the process $V^p(x,\phi , \pA ,\pC)$, and not $V(x,\phi , \pA ,\pC)$, to emphasize the important role of $V^p(x,\phi , \pA ,\pC)$ as the value of the hedger's portfolio of traded assets. Observe also that the equality  $V^p(x,\phi , \pA ,\pC) = V(x,\phi , \pA ,\pC)$ holds when $\pC$ vanishes, that is, $C=0$.

Let the initial endowment of the hedger be denoted by $x$. It is now natural to represent a portfolio as $\phi =
(\xi^1, \dots , \xi^d ,\psi^{l}, \psi^{b},\eta^{b}, \eta^{l})$ with the corresponding wealth process
\bde
V_t(x, \phi , \pA, C ) = \sum_{i=1}^d \xi^i_tS^i_t+ \psi^{l}_t \Blr_t+\psi^{b}_t\Bbr_t+   \etab_t B^{\pCc,b}_t+ \etal_t B^{\pCc,l}_t
\ede
where $\etab_t =-  (B^{\pCc,b}_t)^{-1}\pC_t^+$ and  $\etal_t =(B^{\pCc,l}_t)^{-1} \pC_t^- $ for all $t\in[0,T]$.

\bd
The hedger's trading strategy $(x, \phi , \pA, C )$ is {\it self-financing} whenever the process $V^p(x,\phi , \pA ,\pC)$, which is given by
\be \label{Bergman model value of strategy 1}
V^p_t (x,\phi , \pA ,\pC) = \sum_{i=1}^{d}\xi^i_t S^i_t + \psi^l_t B^l_t+\psi^b_t B^b_t ,
\ee
satisfies
\begin{align*}
V^p_t (x,\phi , \pA ,\pC)  = \, \, &x + \sum_{i=1}^{d} \int_{0}^{t} \xi^i_u \, d(S^i_u + \pA^i_u )
+ \int_0^t \psi^l_u \, dB^l_u + \int_0^t \psi^b_u \, dB^b_u + \pA_t \\
&+ \int_0^t \etab_u\, dB^{\pCc,b}_u + \int_0^t \etal_u \, dB^{\pCc,l}_u - \VCc_t (x,\phi , \pA ,\pC)
\end{align*}
where
\bde
\VCc_t (x,\phi , \pA ,\pC) = \etab_t B^{\pCc,b}_t+ \etal_t B^{\pCc,l}_t=-C_{t}.
\ede
\ed

 We make the natural assumption that $\psi^{l}_t \geq 0$ and $\psi^{b}_t \leq 0$ for all $t \in [0,T]$. Since simultaneous lending and borrowing of cash is either formally precluded or it is sub-optimal (if $\rbb \geq \rll$, as we will postulate in Assumption \ref{assumption for absolutely continuous}), we also postulate that $\psi^{l}_t \psi^{b}_t =0$ for all $t \in [0,T]$. Consequently, using (\ref{Bergman model value of strategy 1}), we obtain the following equalities
\bde
\psi^{l}_t = (\Blr_t)^{-1} \Big( V^p_t (x,\phi , \pA ,\pC) -\sum_{i=1}^d \xi^i_t S^i_t\Big)^+, \quad
\psi^{b}_t = - (\Bbr_t)^{-1} \Big( V^p_t (x,\phi , \pA ,\pC) - \sum_{i=1}^d\xi^i_t S^i_t \Big)^-.
\ede

\bhyp \label{assumption for absolutely continuous}
The collateral accounts $B^{c,l}$ and $B^{c,b}$ satisfy $B^{\pCc,l}=B^{\pCc,b}=B^{c}$ where $B^c$ is absolutely
continuous, so that $dB^{c}_{t}=r^{c}_{t}B^{c}_{t}\, dt$ for some $\mathbb{G}$-adapted process $r^{c}$.
The riskless accounts are absolutely continuous, so that they can be represented as $dB^{l}_{t}=r^{l}_{t}B^{l}_{t}\, dt$
and $dB^{b}_{t}=r^{b}_{t}B^{b}_{t}\, dt$ for some $\mathbb{G}$-adapted processes $r^{l}$ and $r^{b}$ such that $0 \leq \rll \leq \rbb$.
\ehyp

In view of Assumption \ref{assumption for absolutely continuous}, we have
\begin{align} \label{definition for FC}
F^{C}_t&:=\int_0^t \etab_u\, dB^{\pCc,b}_u + \int_0^t \etal_u \, dB^{\pCc,l}_u \nonumber \\
&=-\int_0^t \pC_u^+ (B^{\pCc,b}_u)^{-1}\, dB^{\pCc,b}_u + \int_0^t \pC_u^- (B^{\pCc,l}_u)^{-1}\, dB^{\pCc,l}_u \\
&=-\int_0^t  \pC_u (B^{\pCc}_u)^{-1} \, dB^{\pCc}_u = -\int_0^t  r^c_u \pC_u \, du. \nonumber
\end{align}
For brevity, we will write $A^{C}:=A+C+F^{C}$. Moreover, we introduce the auxiliary processes $\wt S^{i,l,\textrm{cld}}$ and $\wt S^{i,b,\textrm{cld}}$ for $i=1,2, \dots , d$, which are given by the following expressions
\bde
\wt S^{i,l,{\textrm{cld}}}_t :=  (\Blr_t)^{-1}S^i_t + \int_{(0,t]} (\Blr_u)^{-1} \, d\pA^i_u
\ede
and
\bde
\wt
S^{i,b,{\textrm{cld}}}_t ;=  (\Bbr_t)^{-1}S^i_t + \int_{(0,t]} (\Bbr_u)^{-1} \, d\pA^i_u
\ede
so that their dynamics are
\bde 
d\wt S^{i,l,{\textrm{cld}}}_t=(\Blr_t)^{-1}\left(dS^i_t - \rll_t S^i_t \, dt + d\pA^i_t\right)
\ede
and
\bde  
d\wt S^{i,b,{\textrm{cld}}}_t=(\Bbr_t)^{-1}\left(dS^i_t - \rbb_t S^i_t \, dt + d\pA^i_t\right).
\ede
We also denote
\bde
A^{C,l}_t := \int_{(0,t]}(\Blr_{u})^{-1}\, dA^C_{u}, \quad A^{C,b}_t:= \int_{(0,t]}(\Bbr_{u})^{-1}\, dA^C_{u}.
\ede
Under Assumption \ref{assumption for absolutely continuous}, the self-financing condition for the trading strategy $(x,\phi , \pA ,\pC)$ reads
\begin{align*}
dV^p_t (x,\phi , \pA ,\pC)  =  & \sum_{i=1}^d \xi^i_t \, (dS^i_t + d\pA^i_t)+ d\pA_t^{C}
+ r_{t}^{l}\Big(V^p_t (x,\phi , \pA ,\pC) -\sum_{i=1}^d\xi^i_t S^i_t\Big)^+ dt\nonumber\\
&-  r_{t}^{b}\Big( V^p_t (x,\phi , \pA ,\pC)-\sum_{i=1}^d\xi^i_t S^i_t \Big)^- dt .
\end{align*}
This leads to the following proposition whose easy proof is omitted.

\bp \label{pocc}
The process $Y^{l}:=(B^{l})^{-1}V^p(x,\phi , \pA ,\pC)$ satisfies
\be  \label{Bergman model lending BSDE}
dY_{t}^{l} = \sum_{i=1}^dZ^{l,i}_t \, d\wt S^{i,l,{\textrm{cld}}}_t+G_{l}(t,Y_{t}^{l},Z_{t}^{l})\, dt+ d\pA_t^{C,l}
\ee
where  $Z^{l,i}=\xi^i ,\, i=1,2,\ldots,d$ and the mapping $G_l$ equals, for all $(\omega , t,y,z)\in \Omega \times [0,T] \times \mathbb{R}\times\mathbb{R}^{d}$,
\bde
G_{l}(t,y,z)=\sum_{i=1}^d r_{t}^{l}(B_{t}^{l})^{-1} z^i S^i_t+(B_{t}^{l})^{-1}\bigg(r_{t}^{l}\Big(yB_{t}^{l}-\sum_{i=1}^d z^i S^i_t\Big)^+-r_{t}^{b}\Big(yB_{t}^{l}-  \sum_{i=1}^d z^i S^i_t\Big)^- \bigg)-r_{t}^{l}y.
\ede
The process $Y^{b}:=(B^{b})^{-1}V^p (x,\phi , \pA ,\pC)$  satisfies
\bde
dY_{t}^{b} = \sum_{i=1}^dZ^{b,i}_t \, d\wt S^{i,b,{\textrm{cld}}}_t+G_{b}(t,Y_{t}^{b},Z_{t}^{b})\, dt+ d\pA_t^{C,b}
\ede
 where $Z^{b,i}=\xi^i,\, i=1,2,\ldots,d$ and the mapping $G_b$ equals, for all $(\omega ,t,y,z)\in \Omega \times [0,T] \times \mathbb{R}\times\mathbb{R}^{d}$,
\bde
G_{b}(t,y,z)=\sum_{i=1}^d r_{t}^{b}(B_{t}^{b})^{-1}z^i S^i_t+(B_{t}^{b})^{-1}\bigg(r_{t}^{l}\Big(yB_{t}^{b}-\sum_{i=1}^d z^i S^i_t\Big)^+-r_{t}^{b}\Big(yB_{t}^{b}-  \sum_{i=1}^d z^i S^i_t\Big)^- \bigg)-r_{t}^{b}y.
\ede
\ep

The concept of the {\it netted wealth} was introduced in \cite{BR-2014} to study the arbitrage-free property of a model.

\bd \label{nettled wealth}
The {\it netted wealth} $\Vnet(x, \phi , \pA, C)$ of a trading strategy $(x, \phi, \pA, C)$ is given by
$\Vnet(x, \phi , \pA, C):= V(x, \phi , \pA, C) + V(0, \wt \phi  , -\pA, -C)$ where $(0,  \wt \phi ,-A, -C)$
is the unique self-financing strategy satisfying the following conditions: \hfill \break
(i) $V_0(0, \wt \phi  , -\pA ,-C ) = - A_0 $, \hfill \break
(ii) $\widetilde{\xi}^i_t=0$ for all $i=1,2,\dots ,d$ and $t \in [0,T]$, \hfill \break
(iii) $\wt{\psi}^l_t \geq 0 ,\, \wt{\psi}^b_t \leq 0$ and  $\wt{\psi}^l_t \wt{\psi}^b_t =0$ for all $t \in [0,T]$.
\ed

It is worth noting that $\Vnet_0 (x, \phi , \pA, C) =x$ for any contract $(A,C)$ and any strategy $\phi $.
The proof of the next lemma is elementary and thus it is omitted (see Lemma 3.1 in \cite{NR2}).

\bl \label{nettled wealth formula}
We have $\Vnet(x, \phi , \pA, C) =  V(x, \phi , \pA, C) + U(A,C)$,
where the $\gg$-adapted process of finite variation $U(A,C)=U$ is the unique solution to the following equation
\bde
U_t = \int_0^t (\Blr_u)^{-1} ( U_u-C_{u})^+\, d\Blr_u - \int_0^t (\Bbr_u)^{-1} ( U_u-C_{u})^-\, d\Bbr_u - F^{C}_{t} - A_t
\ede
where $F^{C}$ is given by (\ref{definition for FC}). Under Assumption \ref{assumption for absolutely continuous},
we obtain
\bde
U_t = \int_0^t r^l_u( U_u-C_{u})^+\, du - \int_0^t r^b_u ( U_u-C_{u})^-\, du + \int_0^t  r^c_u \pC_u \, du - A_t.
\ede
\el

\subsection{Arbitrage-Free Property}      \label{sect2.3}

Depending on the signs of the initial endowments, we will formally work under two alternative assumptions regarding
a general set-up considered in this work. It is worth noting, however, that these assumptions may in fact be equivalent when a particular model for the dynamics of risky assets is adopted.

\bhyp \label{assumption for lending cumulative dividend price}
There exists a probability measure $\PT^l $ equivalent to $\P $ such that the
processes $\wt S^{i,l,\textrm{cld}},\, i=1,2, \dots ,d$ are $(\PT^l , \gg)$-local martingales.
\ehyp

\bhyp \label{assumption for borrowing cumulative dividend price}
There exists a probability measure $\PT^b $ equivalent to $\P $ such that the
processes $\wt S^{i,b,\textrm{cld}},\, i=1,2, \dots ,d$ are $(\PT^b , \gg)$-local martingales.
\ehyp

In the foregoing definition of admissibility, the discounted netted wealth $\widehat{V}^{net}(x, \phi ,A,C)$ is defined either as
$V^{net}(x, \phi ,A,C)/B^l$, if Assumption \ref{assumption for lending cumulative dividend price} is
postulated, or as $V^{net}(x, \phi ,A,C)/B^b$, when Assumption \ref{assumption for borrowing cumulative dividend price}
is valid. The same notational convention is used in Proposition \ref{Bergman model proposition for arbitrage free}.

\bd \label{arbitrage using nettled wealth}
A self-financing trading strategy $(x,\phi ,A,C)$ is {\it admissible for the hedger} whenever the discounted netted wealth process $\widehat{V}^{net}(x, \phi ,A,C)$ is bounded from below by a constant.
\ed

\bd
An admissible trading strategy $(x, \phi ,A,C)$ is an {\it arbitrage opportunity for the hedger} with respect to $(A,C)$ whenever
\be
\P ( \Vnet_T(x, \phi , A, C) \geq \VLL_T (x))=1\quad \text{ and }\quad \P ( \Vnet_T (x,\phi ,A,C) > \VLL_T (x) ) > 0 \nonumber
\ee
where $\VLL_t (x) := x^+ B^l_t - x^- B^b_t$ for all $t \in [0,T]$.  A market model is said to be {\it  arbitrage-free} for the hedger if there is no arbitrage opportunity for the hedger in regard to any contract $(A,C)$.
\ed

\bp \label{Bergman model proposition for arbitrage free}
We consider the market model introduced in this section under Assumption \ref{assumption for absolutely continuous}. \hfill \break
(i) If Assumption \ref{assumption for lending cumulative dividend price} holds and $x_1 \geq 0 ,\, x_2 \geq 0$,
then the market model is arbitrage-free with respect to any contract $(A,C)$ for the hedger and the counterparty. \hfill \break (ii) If Assumption \ref{assumption for borrowing cumulative dividend price} holds and $x_1 \leq 0 ,\, x_2 \geq 0$, then
the market model is arbitrage-free with respect to any contract $(A,C)$ for the hedger and the counterparty.
\ep

\proof
We only prove the non-arbitrage property of the model from the perspective of hedger with a positive initial endowment $x = x_1 \ge0$, since all other cases can be proven using analogous arguments. From (\ref{Bergman model lending BSDE}) and $\rll \leq \rbb$, we know that
$V^{l} (x, \phi , A,C) =(B^{l})^{-1}V^p (x,\phi , \pA ,\pC)$ satisfies
\begin{align*}
d V^{l}_t (x,\phi , \pA ,C)\leq\sum_{i=1}^d \xi^i_t \, d\wt S^{i,l,{\textrm{cld}}}_t + d\pA_t^{C,l}.
\end{align*}
Furthermore, in view of Lemma \ref{nettled wealth formula}, the netted wealth is given by $\Vnet (x,\phi , \pA ,C) =  V(x,\phi , \pA , C) + U(A,C)$, where in turn the $\gg$-adapted process of finite variation $U(A,C)$ is given by Lemma \ref{nettled wealth formula}. Hence the process
$V^{l,net}(x,\phi, A,C):= (\Blr)^{-1}\Vnet(x ,\phi , \pA ,C)=V^{l}(x,\phi, A,C)+ (\Blr)^{-1}U(A,C)$ satisfies
\bde
dV_t^{l,net} (x,\phi, A,C) \leq\sum_{i=1}^d \xi^i_t \, d\wt S^{i,l,{\textrm{cld}}}_t + (r^l_t - r^b_t )(\Blr_t)^{-1} ( U_t(A,C)-C_{t})^- \, dt
\leq\sum_{i=1}^d \xi^i_t \, d\wt S^{i,l,{\textrm{cld}}}_t
\ede
or, more explicitly,
\be \label{inequality for proving arbitrage free}
(B^l_t)^{-1} \big( V^{net}_t(x,\phi, A,C) - x \big) \leq \sum_{i=1}^d \int_{(0,t]} \xi^i_t \, d\wt S^{i,l,{\textrm{cld}}}_t
\ee
since $\Vnet(x, \phi , \pA, C) =x$. The assumption that the process $V^{l,net}$ is bounded from below, implies that the right-hand side in \eqref{inequality for proving arbitrage free} is a $(\PT^l,\gg)$-supermartingale, which is null at $t=0$. Next, since $x\ge0$, we have that $\VLL_T(x) = \Blr_T x$ and thus, from \eqref{inequality for proving arbitrage free}, we obtain
\bde
(\Blr_T)^{-1} \big( \Vnet_T (x,\phi , A ,C) - \VLL_T (x) \big) \leq \sum_{i=1}^d \int_{(0,T]} \xi^i_t \, d\wt S^{i,l,{\textrm{cld}}}_t.
\ede
Since the probability measure $\PT^l $ was assumed to be equivalent to $\P$, we conclude that either the equality $\Vnet_T (x,\phi , A,C) = \VLL_T(x)$ holds or $\P ( \Vnet_T (x,\phi , A,C) < \VLL_T(x))>0$. This means that arbitrage opportunities are precluded and thus the model is arbitrage-free for the hedger in regard to any contract $(A,C)$.
\endproof

\section{Ex-Dividend Prices and Related Pricing BSDEs}  \label{sect3}

The main goal of this section is to show that, under mild technical assumption, the range of fair
bilateral prices of a generic collateralized contract $(A,C)$ is non-empty for some choices of
initial endowments of the hedger and the counterparty.

\subsection{Generic Market Models} \label{sect3.1}

To show the existence of a solution to the pricing BSDE, we need to complement Assumptions \ref{assumption for lending cumulative dividend price} and \ref{assumption for borrowing cumulative dividend price} by imposing specific conditions on the underlying market model. In essence, we postulate that the discounted cumulative prices of risky assets are continuous martingales under
an equivalent probability measure and its quadratic variation process satisfies suitable technical conditions.

We define the matrix-valued process $\mathbb{S}$
\[
\mathbb{S}_{t}:=
\begin{pmatrix}
S^{1}_{t} & 0 & \ldots & 0 \\
0 & S^{2}_{t} & \ldots & 0 \\
\vdots & \vdots & \ddots & \vdots\\
0 & 0 & \ldots & S^{d}_{t}
\end{pmatrix}.
\]

We will work under the following alternative assumptions regarding the quadratic variation process
for continuous martingales $\wt S^{l,\textrm{cld}}$ and $\wt S^{b,\textrm{cld}}$. Note that $^{\ast}$ stands for
the transposition.

\bhyp \label{changed assumption for lending cumulative dividend price}
We postulate that: \hfill \break
(i) the process $\wt S^{l,\textrm{cld}}$ is a continuous, square-integrable, $(\PT^l , \gg)$-martingale and has the predictable representation property with respect to the filtration $\gg$ under~$\PT^l$, \hfill \break
(ii)  there exists an $\mathbb{R}^{d\times d}$-valued, $\gg$-adapted process $m^{l}$ such that
\be \label{vfvf1}
\langle \wt S^{l,\textrm{cld}}\rangle_{t}=\int_{0}^{t}m^{l}_{u}(m_{u}^{l})^{\ast}\,du
\ee
with the process $m^{l}(m^{l})^{\ast}$ is invertible and satisfies $m^{l}(m^{l})^{\ast}=\mathbb{S}\sigma\sigma^{\ast}\mathbb{S}$ where $\sigma$ is a $d$-dimensional square matrix of $\gg$-adapted processes satisfying the {\it ellipticity condition}: there exists a constant $\Lambda>0$
\be \label{elli}
\sum_{i,j=1}^{d}\left(\sigma_{t}\sigma^{\ast}_{t}\right)_{ij}a_{i}a_{j}\ge \Lambda|a|^{2}=\Lambda a^{\ast}a,\quad \forall \, a\in\mathbb{R}^{d},\, t\in[0,T].
\ee
\ehyp

\bhyp \label{changed assumption for borrowing cumulative dividend price}
We postulate that: \hfill \break
(i) the process $\wt S^{b,\textrm{cld}}$ is a continuous, square-integrable $(\PT^b , \gg)$-martingale
and has the predictable representation property with respect to the filtration $\gg$ under~$\PT^b$, \hfill \break
(ii) there exists an $\mathbb{R}^{d\times d}$-valued, $\gg$-adapted process $m^{b}$ such that
\be \label{vfvf2}
\langle \wt S^{b,\textrm{cld}}\rangle_{t}=\int_{0}^{t}m^{b}_{u}(m_{u}^{b})^{\ast}\,du
\ee
with the process $m^{b}(m^{b})^{\ast}$ is invertible and satisfies $m^{b}(m^{b})^{\ast}= \mathbb{S}\sigma\sigma^{\ast}\mathbb{S}$
where $\sigma$ is a $d$-dimensional square matrix of $\gg$-adapted processes satisfying the ellipticity condition \eqref{elli}.
\ehyp

\subsection{Prices and Hedging Strategies} \label{sect3.2}

Definition of the ex-dividend price for the hedger and the counterparty is based on replication of all
cash flows associated with a given contract $(A,C)$.

\bd \lab{def:replicate}
For a fixed $t \in [0,T]$, a self-financing trading strategy $(\VLL_{t}(x)+p_t, \phi , A- A_t , \pC )$,
where $p_t$ is a ${\cal G}_{t}$-measurable random variable, is said to {\it replicate the collateralized
contract} $(A,C)$ on $[t,T]$ whenever $V_T(\VLL_{t}(x)+p_t , \phi , A-A_t ,C) = \VLL_T (x)$.
\ed

 Since we deal here with a non-linear pricing rule, we need to examine
separately the pricing problem for each party and take into account their initial endowments.
Of course, if we postulate that we work within a linear framework in which all interest rates coincide, that is, $r^l =r^b = r^c$,
then, as expected, we obtain the equality $P^{h}_t(x_1, A, C) = P^{c}_t(x_1, A, C)$ for every contract $(A,C)$ and all $t$.

\bd \label{definition of ex-dividend price}
Any ${\cal G}_{t}$-measurable random variable for which a replicating strategy for $(A,C)$ over $[t,T]$ exists is called the {\it hedger's ex-dividend price} at time $t$ for a contract $(A,C)$ and it is denoted by $P^{h}_t(x_1, A, C)$, so that for some
$\phi $ replicating $(A,C)$
\bde
V_T(\VLL_{t}(x_1)+P^{h}_t(x_1, A, C), \phi , A -A_t, C) = \VLL_T (x_1).
\ede
For an arbitrary level $x_2$ of the counterparty's initial endowment and a strategy $\widetilde{\phi }$ replicating $(-A,-C)$,
the {\it counterparty's ex-dividend price} $P^{c}_t(x_2, -A, -C)$ at time $t$ for a contract $(-A,-C)$ is implicitly given by the equality
\bde
V_T(\VLL_{t}(x_2)-P^{c}_t(x_2, -A, -C), \widetilde{ \phi }, -A+A_t, -C) = \VLL_T (x_2).
\ede
\ed

By a {\it fair bilateral price}, we mean the price level at which no arbitrage opportunity arises for either party.
Hence the range of fair bilateral prices at time $t$ is defined as follows.

\bd  \label{range of fair}
The  $\G_t$-measurable interval
\bde
{\cal R}^f_t (x_1,x_2) := \big[ P^{c}_t(x_{2},-A,-C), P^{h}_t (x_{1},A,C) \big]
\ede
is called the {\it range of fair bilateral prices} at time $t$ of an OTC contract $(A,C)$ between the hedger and the counterparty.
\ed

We are in a position to state the results furnishing the ex-dividend prices and replicating strategies for the hedger and
the counterparty. Their proofs hinge on a combination of results on BSDEs from \cite{NR3} with arguments used in \cite{BR-2014}.
It is worth noting that in Propositions \ref{Bergman model hedger ex-dividend price} and \ref{Bergman model counterparty ex-dividend price}, the pricing BSDE is driven either by the process $\wt S^{l,{\textrm{cld}}}$ or the process
$\wt S^{b,{\textrm{cld}}}$, depending on whether the initial endowment is positive or negative. This is somewhat inconvenient
when we wish to compare prices for both parties, and thus we will also derive in Proposition \ref{Bergman model general pricing proposition} pricing BSDEs driven by a common process, denoted by $\wt S^{{\textrm{cld}}}$. It is fair to acknowledge, however, that the financial interpretation of the auxiliary process $\wt S^{{\textrm{cld}}}$ is not as transparent as that of the discounted cumulative prices $\wt S^{l,{\textrm{cld}}}$ and $\wt S^{b,{\textrm{cld}}}$, and thus the  process should be seen as a purely mathematical artifact.

Following \cite{NR3}, but with $Q_t=t$, we denote by $\wHzerd $ the subspace of all $\mathbb{R}^{d}$-valued, $\gg$-adapted processes $X$ with
\be \label{defhh}
|X|_{\wHzerd}^{2}:=\EP \bigg[ \int_{0}^{T}\|X_{t}\|^{2}\,dt \bigg] <\infty .
\ee
Also, let $\widehat{L}^{2}_{0}$ stand for the space of all real-valued, $\mathcal{G}_{T}$-measurable
random variables $\eta$ such that $|\eta|_{\widehat{L}^2_{0}}^{2}=\EP (\eta^{2})<\infty $.

\bd
A contract $(A,C)$ is {\it admissible under} $\PT^l$ if the process $A^{C,l}$ belongs to $\wHzero$
and the random variable $A^{C,l}_{T}$ belongs to $\widehat{L}^{2}_{0}$ under~$\PT^l$.
A contract $(A,C)$ is {\it admissible under} $\PT^b$ if the process $A^{C,b}$ belongs to $\wHzero$
and the random variable $A^{C,b}_{T}$ belongs to $\widehat{L}^{2}_{0}$ under~$\PT^b$.
\ed

From now on, we postulate that the processes $r^{l}$ and $r^{b}$ are nonnegative and bounded.

\bp \label{Bergman model hedger ex-dividend price}
(i) Let the hedger's initial endowment $x_1 = x \geq 0$ and let Assumption \ref{changed assumption for lending cumulative dividend price} be satisfied. Then for any contract $(A,C)$ admissible under $\PT^l$, the hedger's ex-dividend price equals $P^{h}(x,A,C) =  \Blr (Y^{h,l,x} - x)-C$ where $(Y^{h,l,x}, Z^{h,l,x})$ is the unique solution to the BSDE
\begin{equation} \label{Bergman model BSDE with positive x for hedger}
\left\{ \begin{array}
[c]{l}
dY^{h,l,x}_t = Z^{h,l,x,\ast}_t \, d \wt S^{l,{\textrm{cld}}}_t
+G_l \big(t, Y^{h,l,x}_t, Z^{h,l,x}_t \big)\, dt + dA^{C,l}_t, \medskip\\
Y^{h,l,x}_T=x.
\end{array} \right.
\end{equation}
The unique replicating strategy for the hedger equals $\phi = \big(\xi^1,\dots ,\xi^d, \psi^{l}, \psi^{b}, \etab, \etal\big)$ where for all $t\in[0,T]$ and $i=1,2,\ldots,d$
\bde
\xi^i_{t}= Z^{h,l,x,i}_{t},  \quad
\etab_t =-  (B^{\pCc,b}_t)^{-1}\pC_t^+, \quad
\etal_t =(B^{\pCc,l}_t)^{-1} \pC_t^- ,
\ede
and
\bde
\psi^{l}_t = (\Blr_t)^{-1} \Big( \Blr_tY^{h,l,x}_{t} -\sumik_{i=1}^d\xi^i_t S^i_t \Big)^+, \quad
\psi^{b}_t = - (\Bbr_t)^{-1} \Big(\Blr_tY^{h,l,x}_{t}-\sumik_{i=1}^d \xi^i_t S^i_t\Big)^-.
\ede
(ii)  Let the hedger's initial endowment $x_1 = x \leq 0$ and let Assumption \ref{changed assumption for borrowing cumulative dividend price} be satisfied. Then for any contract $(A,C)$ admissible under $\PT^b$, the hedger's ex-dividend price equals $P^{h}(x,A,C) =  \Bbr (Y^{h,b,x} - x)- C$  where $(Y^{h,b,x}, Z^{h,b,x})$ is the unique solution to the BSDE
\begin{equation}\label{Bergman model BSDE with negative x for hedger}
\left\{ \begin{array}
[c]{l}
dY^{h,b,x}_t = Z^{h,b,x,\ast}_t \, d \wt S^{b,{\textrm{cld}}}_t
+G_b \big(t, Y^{h,b,x}_t, Z^{h,b,x}_t \big)\, dt + dA^{C,b}_t, \medskip\\
Y^{h,b,x}_T=x.
\end{array} \right.
\end{equation}
The unique replicating strategy for the hedger equals $\phi = \big(\xi^1,\dots ,\xi^d, \psi^{l}, \psi^{b},\etab, \etal\big)$ where for all $t\in[0,T]$ and $i=1,2,\ldots,d$
\bde
\xi^i_{t}= Z^{h,b,x,i}_{t},  \quad
\etab_t =-  (B^{\pCc,b}_t)^{-1}\pC_t^+, \quad
\etal_t =(B^{\pCc,l}_t)^{-1} \pC_t^-,
\ede
and
\bde
\psi^{l}_t = (\Blr_t)^{-1} \Big( \Bbr_tY^{h,l,x}_{t} -\sumik_{i=1}^d \xi^i_t S^i_t\Big)^+, \quad
\psi^{b}_t = - (\Bbr_t)^{-1} \Big(\Bbr_tY^{h,l,x}_{t}-\sumik_{i=1}^d \xi^i_t S^i_t\Big)^-.
\ede
\ep

\proof
From Theorem 4.1 in \cite{NR3}, we know that if Assumption \ref{changed assumption for lending cumulative dividend price} holds, then BSDE (\ref{Bergman model BSDE with positive x for hedger}) has a unique solution $(Y^{h,l,x}, Z^{h,l,x})$. As in the proof of Proposition 5.2 in \cite{BR-2014}, we can show that $P^{h}(x,A,C) =  \Blr (Y^{h,l,x} - x)$ and derive the unique replicating strategy $\varphi$.
\endproof

\bp \label{Bergman model counterparty ex-dividend price}
For any value $x =x_2$ of the initial endowment, the counterparty's ex-dividend price equals
\bde
P^{c} (x,-A,-C) =-\left(\Blr (Y^{c,l,x} - x)+C\right)\I_{\{x\ge0\}}-\left(\Bbr (Y^{c,b,x} - x)+C\right)\I_{\{x\leq0\}}
\ede
where $(Y^{c,l,x}, Z^{c,l,x})$ and $(Y^{c,b,x}, Z^{c,b,x})$ are respectively the unique solutions to the BSDEs
\bde
\left\{ \begin{array}
[c]{l}
dY^{c,l,x}_t = Z^{c,l,x,\ast}_t \, d \wt S^{l,{\textrm{cld}}}_t
+G_l \big(t, Y^{c,l,x}_t, Z^{c,l,x}_t \big)\, dt -dA^{C,l}_t, \medskip\\
Y^{c,l,x}_T=x , \end{array}
\right.
\ede
and
\bde
\left\{ \begin{array}
[c]{l}
dY^{c,b,x}_t = Z^{c,b,x,\ast}_t \, d \wt S^{b,{\textrm{cld}}}_t
+G_b \big(t, Y^{c,b,x}_t, Z^{c,b,x}_t \big)\, dt -dA^{C,b}_t, \medskip\\
Y^{c,b,x}_T=x. \end{array}
\right.
\ede
The unique replicating strategy for the counterparty equals $\phi = \big(\xi^1,\dots ,\xi^d, \psi^{l}, \psi^{b},\etab, \etal\big)$
where for all $t\in[0,T]$ and $i=1,2,\ldots,d$
\bde
\xi^i_{t}= Z^{c,l,x}_{t}\I_{\{x\ge0\}}+Z^{c,b,x}_{t}\I_{\{x\leq0\}},\quad
\etab_t =-  (B^{\pCc,b}_t)^{-1}\pC_t^-, \quad
\etal_t =(B^{\pCc,l}_t)^{-1} \pC_t^+,
\ede
and
\bde
\begin{array} [c]{ll}
\psi^{l}_t = (\Blr_t)^{-1} \Big( \Blr_tY^{c,l,x}_{t}\I_{\{x\ge0\}}+\Bbr_tY^{c,b,x}_{t}\I_{\{x\leq0\}}-\sum_{i=1}^d\xi^i_t S^i_t \Big)^+, \medskip\\
\psi^{b}_t = - (\Bbr_t)^{-1} \Big(\Blr_tY^{c,l,x}_{t}\I_{\{x\ge0\}}+\Bbr_tY^{c,b,x}_{t}\I_{\{x\leq0\}}-\sum_{i=1}^d\xi^i_t S^i_t \Big)^-. \end{array}
\ede
\ep

\proof
The proof of Proposition \ref{Bergman model counterparty ex-dividend price} is analogous
to the proof of Proposition \ref{Bergman model hedger ex-dividend price} and thus it is omitted.
\endproof

In order to establish the comparison result for ex-dividend prices when the two parties have arbitrary initial endowments, we need a result when the prices are given by solution to two BSDEs driven by the same continuous martingale. To this end, we introduce the following assumption about the underlying financial model.

\bhyp \label{changed assumption for artifical cumulative dividend price}
We postulate that: \hfill \break
(i) there exists a probability measure $\PTb $ equivalent to $\P$ such that the processes $\wt S^{i,\textrm{cld}},\, i=1,2, \dots ,d$ given by (\ref{auxiliary processes})
\be\label{auxiliary processes}
d\wt S^{i,\textrm{cld}}_t = dS^i_t + d\pA^i_t - \beta^{i}_{t}S_{t}^{i}\,dt
\ee
for some $\gg$-adapted bounded processes $\beta^{i}$ satisfying $r^{b}\leq\beta^{i}$, are $(\PTb , \gg)$-continuous square-integrable martingales and have the predictable representation property with respect to the filtration $\gg$ under $\PTb$, \hfill \break
(ii) there exists an $\mathbb{R}^{d\times d}$-valued, $\gg$-adapted process $m$ such that
\be\label{auxiliary processes quadratic variation}
\langle \wt S^{\textrm{cld}}\rangle_{t}=\int_{0}^{t}m_{u}m_{u}^{\ast}\,du
\ee
where $m m^{\ast}$ is invertible and satisfies $m m^{\ast}= \mathbb{S}\sigma\sigma^{\ast}\mathbb{S}$ where a $d$-dimensional square matrix $\sigma$ of $\gg$-adapted processes satisfies the ellipticity condition
\eqref{elli}.
\ehyp

\bd
We say that $(A,C)$ is {\it admissible under} $\PTb$ when $A^C \in\wHzero $ and $A^C_T\in \widehat{L}^{2}_{0}$ under $\PTb$.
\ed

The next result expresses the unilateral prices of a contract $(A,C)$ is terms of solutions to BSDEs driven by the continuous $\PTb$-martingale $\wt S^{\textrm{cld}}$. It will be used in the next section to study the range of fair bilateral prices. To alleviate notation, we denote
\bde
G(t,y,z)=\rll_t \Big(y- \sumik_{i=1}^dz^{i}S^i_t\Big)^+- \rbb_t \Big( y-\sumik_{i=1}^d z^{i}S^i_t \Big)^-.
\ede

\bp \label{Bergman model general pricing proposition}
Let Assumption \ref{changed assumption for artifical cumulative dividend price} be valid. Then for any $x_{1},x_{2}\in\mathbb{R}$ and an arbitrary contract $(A,C)$ admissible under $\PTb$, we have that $P^{h}(x_{1},A,C)=\widetilde{Y}^{h,x_{1}}-C$ and $P^{c}(x_{2},-A,-C)=\widetilde{Y}^{c,x_{2}}-C$ where  $(\widetilde{Y}^{h,x_{1}},\widetilde{Z}^{h,x_{1}})$ is the unique solution
to the following BSDE
\bde
\left\{ \begin{array}
[c]{ll}
d\widetilde{Y}^{h,x_{1}}_t =\widetilde{Z}^{h,x_{1},\ast}_t\, d\wt S_t^{\textrm{cld}}+G^{h}(t,x_{1}, \widetilde{Y}^{h,x_{1}}_t,\widetilde{Z}^{h,x_{1}}_t)\,dt+dA^C_t,\medskip\\
\widetilde{Y}^{h,x_{1}}_T=0,
\end{array} \right.
\ede
and $(\widetilde{Y}^{c,x_{2}},\widetilde{Z}^{c,x_{2}})$ is the unique solution to the following BSDE
\bde
\left\{ \begin{array}
[c]{ll}
d\widetilde{Y}^{c,x_{2}}_t=\widetilde{Z}^{c,x_{2},\ast}_t\, d\wt S_t^{\textrm{cld}}+G^{c}(t, x_{2}, \widetilde{Y}^{c,x_{2}}_t,\widetilde{Z}^{c,x_{2}}_t)\,dt+dA^C_t,\medskip\\
\widetilde{Y}^{c,x_{2}}_T=0,
\end{array} \right.
\ede
where
\bde
G^{h}(t,x,y,z):=\sum_{i=1}^dz^{i}\beta^{i}_{t}S_{t}^{i}+\big(-x\rll_t\Blr_t+G(t,y+x\Blr_t,z)\big)\I_{\{x\ge0\}}
+ \big( -x\rbb_t\Bbr_t+G(t, y+x\Bbr_t,z)\big) \I_{\{x\leq0\}}
\ede
and
\bde
G^{c}(t,x,y,z):=\sum_{i=1}^dz^{i}\beta^{i}_{t}S_{t}^{i}+ \big( x\rll_t\Blr_t-G(t,-y+x\Blr_t,-z)\big) \I_{\{x\ge0\}} + \big( x\rbb_t\Bbr_t-G(t, -y+x\Bbr_t,-z) \big )\I_{\{x\leq0\}}.
\ede
The unique replicating strategy for the hedger equals $\phi = \big(\xi^1,\dots ,\xi^d, \psi^{l}, \psi^{b}, \etab, \etal\big)$ where for all $t\in[0,T]$ and $i=1,2,\ldots,d$
\bde
\xi^i_{t}= \widetilde{Z}^{h,x_{1},i}_{t}, \quad \etab_t =-  (B^{\pCc,b}_t)^{-1}\pC_t^+, \quad \etal_t =(B^{\pCc,l}_t)^{-1} \pC_t^-,
\ede
and
\bde
\begin{array} [c]{ll}
\psi^{l}_t = (\Blr_t)^{-1} \Big( \widetilde{Y}^{h,x_{1}}_{t}+x_{1}\Blr_t\I_{\{x_{1}\ge0\}}+x_{1}\Bbr_t\I_{\{x_{1}\leq0\}}- \sum_{i=1}^d\xi^i_t S^i_t \Big)^+, \medskip\\
\psi^{b}_t = - (\Bbr_t)^{-1} \Big( \widetilde{Y}^{h,x_{1}}_{t}+x_{1}\Blr_t\I_{\{x_{1}\ge0\}}+x_{1}\Bbr_t\I_{\{x_{1}\leq0\}}- \sum_{i=1}^d\xi^i_t S^i_t \Big)^-.
\end{array}
\ede
The unique replicating strategy for the counterparty equals $\phi = \big(\xi^1,\dots ,\xi^d, \psi^{l}, \psi^{b}, \etab, \etal\big)$
where for all $t\in[0,T]$ and $i=1,2,\ldots,d$
\bde
\xi^i_{t}=-\widetilde{Z}^{c,x_{2},i}_{t}, \quad \etab_t =-  (B^{\pCc,b}_t)^{-1}\pC_t^-, \quad \etal_t =(B^{\pCc,l}_t)^{-1} \pC_t^+,
\ede
and
\bde
\begin{array} [c]{ll}
\psi^{l}_t = (\Blr_t)^{-1} \Big(-\widetilde{Y}^{c,x_{2}}_{t}+x_{2}\Blr_t\I_{\{x_{2}\ge0\}}+x_{2}\Bbr_t\I_{\{x_{2}\leq0\}}- \sum_{i=1}^d\xi^i_t S^i_t \Big)^+, \medskip\\
\psi^{b}_t = - (\Bbr_t)^{-1} \Big( -\widetilde{Y}^{c,x_{2}}_{t}+x_{2}\Blr_t\I_{\{x_{2}\ge0\}}+x_{2}\Bbr_t\I_{\{x_{2}\leq0\}}- \sum_{i=1}^d\xi^i_t S^i_t \Big)^-.
\end{array}
\ede
\ep

\subsection{Range of Fair Bilateral Prices}  \label{sect3.3}

We are now in a position to study the range of fair bilateral prices at time $t$ (see Definition \ref{range of fair}).
It appears that, under suitable assumptions, it is non-empty when the initial endowments of the two parties have
the same sign but, in general, it may be empty if the signs are different, that is, when
$x_{1}< 0$ and $x_{2}>0$.

We first examine the case where the initial endowments satisfy $x_{1}\ge0$ and $x_{2}\ge0$.

\bp \label{Bergman model inequality proposition for both positive initial wealth}
Let Assumption \ref{changed assumption for lending cumulative dividend price} be valid.
Then for any $x_{1}\ge0,\, x_{2}\ge0$ and an arbitrary contract $(A,C)$ admissible under $\PT^l$ we have, for every $t\in[0,T]$,
\bde
P^{c}_t (x_{2},-A,-C)\leq P^{h}_t (x_{1},A,C),  \quad \PT^l-\aass ,
\ede
so that the range of fair bilateral prices ${\cal R}^f_t(x_1,x_2)$ is non-empty almost surely.
\ep

\begin{proof}
We assume that $x_{1}\ge0,\, x_{2}\ge0$ and we denote $\bar{Y}^{h,l,x_{1}}:=Y^{h,l,x_{1}} - x_{1}$ and $\bar{Z}^{h,l,x_{1}}=Z^{h,l,x_{1}}$. In view of Propositions  \ref{Bergman model hedger ex-dividend price} and \ref{Bergman model counterparty ex-dividend price}, the pair $(\bar{Y}^{h,l,x_{1}},\bar{Z}^{h,l,x_{1}})$ is the unique solution of the following BSDE
\bde
\left\{\begin{array} [c]{l}
d\bar{Y}^{h,l,x_{1}}_t = \bar{Z}^{h,l,x_{1},\ast}_t \, d \wt S^{l,{\textrm{cld}}}_t
+G_l \big(t, \bar{Y}^{h,l,x_{1}}_t+x_{1}, \bar{Z}^{h,l,x_{1}}_t \big)\, dt + dA^{C,l}_t, \medskip\\
\bar{Y}^{h,l,x_{1}}_T=0.
\end{array}
\right.
\ede
Similarly, $(\bar{Y}^{c,l,x_{2}}, \bar{Z}^{c,l,x_{2}}):=(-(Y^{c,l,x_{2}}- x_{2}),\, \bar{Z}^{c,l,x_{2}}=-Z^{c,l,x_{2}})$ is the unique solution of the following BSDE
\bde
\left\{\begin{array} [c]{l}
d\bar{Y}^{c,l,x_{2}}_t = \bar{Z}^{c,l,x_{2},\ast}_t \, d \wt S^{l,{\textrm{cld}}}_t
-G_l \big(t, -\bar{Y}^{c,l,x_{2}}_t+x_{2}, -\bar{Z}^{c,l,x_{2}}_t \big)\, dt +dA^{C,l}_t, \medskip\\
\bar{Y}^{c,l,x_{2}}_T=0.
\end{array}
\right.
\ede
In view of the comparison theorem for BSDEs (see Theorem 3.3 in \cite{NR3}), if we show that $-G_l \big(t, y+x_{1}, z \big)
\ge G_l \big(t, -y+x_{2}, -z \big)$ for all $(y,z)\in \mathbb{R}\times\mathbb{R}^{d},\, \PT^l\otimes \Leb-\aaee$,
then we will deduce that $\bar{Y}^{h,l,x_{1}}\ge\bar{Y}^{c,l,x_{2}}$. We denote
\begin{align*}
\delta &:=G_l \big(t, y+x_{1}, z \big)+G_l \big(t, -y+x_{2}, -z \big) \\
&=- \rll_t (x_{1}+x_{2})+(\Blr_t)^{-1}\rll_t (\delta_{1}^{+}+\delta_{2}^{+})-(\Blr_t)^{-1}\rbb_t (\delta_{1}^{-}+\delta_{2}^{-})
\end{align*}
where
\bde
\delta_{1}:=\Blr_t y+\Blr_t x_{1}-\sumik_{i=1}^dz^i S^i_t, \quad\delta_{2}:=-\Blr_t y+\Blr_t x_{2}+\sumik_{i=1}^d z^i S^i_t.
\ede
Since $r^l\leq r^b$, we have
\begin{align*}
\delta &=- \rll_t (x_{1}+x_{2})+(\Blr_t)^{-1}\rll_t (\delta_{1}^{+}+\delta_{2}^{+})-(\Blr_t)^{-1}\rbb_t (\delta_{1}^{-}+\delta_{2}^{-})\\
&\leq - \rll_t (x_{1}+x_{2})+(\Blr_t)^{-1}\rll_t (\delta_{1}+\delta_{2})= 0.
\end{align*}
Consequently, we have $\delta\leq 0$, which yields  $-G_l \big(t, y+x_{1}, z \big)
\ge G_l \big(t, -y+x_{2}, -z \big)$, the proof is complete.
\end{proof}

\bp \label{Bergman model inequality proposition for both negative initial wealth}
Let Assumption \ref{changed assumption for borrowing cumulative dividend price} be valid.
Then for any $x_{1}\le0,\, x_{2}\le0$ and an arbitrary contract $(A,C)$ admissible under $\PT^b$ we have, for all $t\in[0,T]$,
\bde
P^{c}_t (x_{2},-A,-C)\leq P^{h}_t (x_{1},A,C),  \quad \PT^b-\aass ,
\ede
so that the range of fair bilateral prices ${\cal R}^f_t (x_1,x_2)$ is non-empty almost surely.
\ep

\begin{proof}
It is now sufficient to show
\bde
-G_b \big(t, y+x_{1}, z \big)
\ge G_b \big(t, -y+x_{2}, -z \big),\ \  \forall \, (y,z)\in \mathbb{R}\times\mathbb{R}^{d}, \ \PT^b \otimes \Leb-\aaee
\ede
If we denote
\bde
\delta :=G_b \big(t, y+x_{1}, z \big)+G_b \big(t, -y+x_{2}, -z \big),
\ede
then, using similar arguments as in the proof of Proposition \ref{Bergman model inequality proposition for both positive initial wealth}, we can prove that $\delta\leq0$.
\end{proof}

Now we consider the case when the initial endowments satisfy $x_{1}\ge0$ and $x_{2}\leq0$.

\bp \label{Bergman model inequality proposition for positive negative initial wealth}
Let Assumption \ref{changed assumption for artifical cumulative dividend price} hold and the initial endowments
satisfy $x_{1}\ge0,\, x_{2}\leq0$. Then the following statements are valid. \hfill \break
(i) If $x_{1}x_{2}=0$, then for any contract $(A,C)$ admissible under $\PTb$ and every $t\in[0,T]$
\be \label{eqnew1}
P^{c}_t (x_{2},-A,-C)\leq P^{h}_t (x_{1},A,C),  \quad \PTb-\aass ,
\ee
so that the range of fair bilateral prices ${\cal R}^f_t (x_1,x_2)$ is non-empty almost surely. \hfill \break
(ii) Assume that $r^{l}$ and $r^{b}$ are deterministic and satisfy $r^{l}_{t}<r^{b}_{t}$ for all $t\in[0,T]$.
Then inequality \eqref{eqnew1} holds for all contracts $(A,C)$ admissible under $\PTb$ and all $t\in[0,T]$
if and only if $x_{1}x_{2}=0$.
\ep

\begin{proof}
(i) If $x_{1}\ge0,\, x_{2}\leq0$, then we can show that
\begin{align*}
\delta&:=g(t,y+x_{1}B_{t}^{l},z)+g(t,-y+x_{2}B_{t}^{b},-z)-x_{1}r_{t}^{l}B_{t}^{l}-x_{2}r_{t}^{b}B_{t}^{b} \\
&\leq \min \, \big\{(\rll_t- \rbb_t )x_{2}B_{t}^{b} , (\rbb_t- \rll_t)x_{1}B_{t}^{l}\big\}.
\end{align*}
Indeed, we have
\bde
\delta = -x_{1}r_{t}^{l}B_{t}^{l}-x_{2}r_{t}^{b}B_{t}^{b} + \rll_t (\delta_{1}^{+}+\delta_{2}^{+}) -
\rbb_t (\delta_{1}^{-}+\delta_{2}^{-}),
\ede
where
\bde
\delta_{1}=y+x_{1}B_{t}^{l}-\sumik_{i=1}^{d}z^{i}S_{t}^{i},\quad \delta_{2}=-y+x_{2}B_{t}^{b}+\sumik_{i=1}^{d}z^{i}S_{t}^{i}.
\ede
From $r^l\leq r^b$, we obtain
\begin{align*}
\delta&:= -x_{1}r_{t}^{l}B_{t}^{l}-x_{2}r_{t}^{b}B_{t}^{b} + \rll_t (\delta_{1}^{+}+\delta_{2}^{+}) -
\rbb_t (\delta_{1}^{-}+\delta_{2}^{-}) \\
&\leq  -x_{1}r_{t}^{l}B_{t}^{l}-x_{2}r_{t}^{b}B_{t}^{b} +\min \, \big\{\rll_t (\delta_{1}+\delta_{2}) ,\rbb_t (\delta_{1}+\delta_{2}) \big\}\\
&= -x_{1}r_{t}^{l}B_{t}^{l}-x_{2}r_{t}^{b}B_{t}^{b} +\min \, \big\{\rll_t (x_{1}B_{t}^{l}+x_{2}B_{t}^{b} ) ,\rbb_t (x_{1}B_{t}^{l}+x_{2}B_{t}^{b} )\big\}\\
&= \min \, \big\{(\rll_t- \rbb_t )x_{2}B_{t}^{b} , (\rbb_t- \rll_t)x_{1}B_{t}^{l}\big\}.
\end{align*}

If $x_{1}x_{2}=0$, then the right-hand side of the above inequality is non-positive. Therefore, $\delta\leq0$ and thus for any contract $(A,C)$ admissible under $\PTb$, from the comparison theorem for BSDEs and Proposition \ref{Bergman model general pricing proposition}, we deduce that for every $t\in[0,T]$
\bde
P^{c}_t (x_{2},-A,-C)\leq P^{h}_t (x_{1},A,C),  \quad \PTb-\aass
\ede
\noindent (ii) We now assume that the interest rates $r^{l}$ and $r^{b}$ are deterministic and satisfy $r^{l}_{t}<r^{b}_{t}$ for all $t\in[0,T]$. If $x_{1}x_{2} \neq 0$, then the example examined in the proof of Proposition 5.4 in \cite{NR2} gives a contract $(A,C)$, such that the inequality
\bde
P^{c}_{0} (x_{2},-A,-C) >  P^{h}_{0} (x_{1},A,C),  \quad \PTb-\aass
\ede
holds in the present framework, so that ${\cal R}^p_0(x_1,x_2)$ is non-empty almost surely.
\end{proof}

The last result of this subsection deals with the case where $x_{1}\leq0$ and $x_{2}\ge0$.

\bp \label{Bergman model inequality proposition for negative positive initial wealth}
Let Assumption \ref{changed assumption for artifical cumulative dividend price} be valid and the initial endowments
satisfy $x_{1}\leq0,\, x_{2}\ge0$. Then the following statements are valid. \hfill \break
(i) If $x_{1}x_{2}=0$, then for every contract $(A,C)$ admissible under $\PTb$ and all $t\in[0,T]$
\be \label{eqnew2}
P^{c}_t (x_{2},-A,-C)\leq P^{h}_t (x_{1},A,C),  \quad \PTb-\aass ,
\ee
so that the range of fair bilateral prices ${\cal R}^f_t (x_1,x_2)$ is non-empty almost surely. \hfill \break
(ii) Assume that $r^{l}$ and $r^{b}$ are deterministic and satisfy $r^{l}_{t}<r^{b}_{t}$ for all $t\in[0,T]$.
Then  inequality \eqref{eqnew2} holds for all contracts $(A,C)$ admissible under $\PTb$ and all $t\in[0,T]$ 
if and only if $x_{1}x_{2}=0$.
\ep

\section{European Claims and Related Pricing PDEs}   \label{sect4}

To alleviate notation,  we assume that $d=1$, so that there is only one risky asset $S=S^1$. This is not a serious restriction,
however, since all results obtained in this subsection can be easily extended to the multi-asset framework. Moreover, we postulate that the interest rates $r^{l}$ and $r^{b}$ are deterministic and thus the only source of randomness is the Brownian
motion appearing in dynamics \eqref{Partial netting model stock price} of the risky asset.

For conciseness, we focus here on the valuation and hedging of an uncollateralized European contingent claim, that is, we set $C=0$. A generic path-independent claim of European style pays a single cash flow $H(S_{T})$ on the expiration date $T>0$, so that
\bde
A_t - A_0 = -H(S_{T})\I_{[T,T]}(t).
\ede
Since we deal here with a Markovian set-up, it is convenient to consider the pricing problem for a contract
initiated at a fixed, but otherwise arbitrary, date $t\in[0,T]$. For any fixed $t <T$, the risky asset $S$ has the ex-dividend price dynamics under $\P$ given by the following expression, for $u \in [t,T]$,
\be \label{Partial netting model stock price}
dS_u = \mu(u,S_{u})\, du +\sigma(u,S_{u})\, dW_u , \quad S_t=\ssx \in \mathcal{O},
\ee
where $W$ is a one-dimensional Brownian motion and $\mathcal{O}$ is the domain of real values that are attainable by the diffusion process $S$ (usually $\mathcal{O}=\mathbb{R}_{+}$). Moreover, the  coefficients $\mu$ and $\sigma$ are such that SDE (\ref{Partial netting model stock price}) has a unique strong solution. We also assume that the volatility coefficient $\sigma$ is bounded and bounded away from zero. Finally, the dividend process equals $\pA^1_t = \int_0^t \kappa( u, S_u) \, du $.

Our first goal is to derive the hedger's pricing PDE for a path-independent European claim. We observe that
\bde
d\wt S^{\textrm{cld}}_u =dS_u + d\pA^1_u -\beta(u, S_{u})\, du=
\big( \mu(u, S_{u})+\kappa(u, S_{u})-\beta(u, S_{u}) \big) du + \sigma (u, S_{u})\, dW_u .
\ede
From the Girsanov theorem, if we denote
\bde
a_{u}:=(\sigma(u, S_{u}))^{-1}\big( \mu(u, S_{u})+\kappa(u, S_{u})-\beta(u, S_{u})\big)
\ede
and define the probability measure $\PTb$ as
\bde
\frac{d\PTb}{d\P}=\exp\left\{-\int_{t}^{T}a_{u}\, dW_{u}
-\frac{1}{2}\int_{t}^{T}|a_{u}|^{2}\, du \right\},
\ede
then $\PTb $ is equivalent to $\P$ and the process $\widetilde{W}$ is the Brownian motion under $\PTb$, where
$d\widetilde{W}_{u}:=dW_{u}+a_{u}\, du$. It is easy to see that
\bde
d\wt S^{\textrm{cld}}_{u}=\sigma (u, S_{u})\, d\widetilde{W}_u
\ede
and thus we conclude that $\wt S^{\textrm{cld}}$ is a $(\PTb , \gg)$-martingale and $\langle \wt S^{\textrm{cld}}\rangle_{u}=\int_{t}^{u}|\sigma(v, S_{v})|^{2}\, dv$. Therefore, Assumption \ref{changed assumption for artifical cumulative dividend price} holds, provided that we assume that the Brownian motion $\widetilde{W}$ has the predictable representation property under $(\gg,\PTb)$. Of course, the latter assumption is not restrictive in the present setup.

We now consider path-independent claims of European style with the unique cash flow at time $T$ given as $H(S_T)$.
From Proposition \ref{Bergman model general pricing proposition}, for any $x_{1}\in\mathbb{R}$ we have
$P^{h}(x_{1},A,C)=\widetilde{Y}^{h,x_{1}}$ where $(\widetilde{Y}^{h,x_{1}},\widetilde{Z}^{h,x_{1}})$ is the unique solution of following BSDE driven by the Brownian motion $\widetilde{W}$
\be \label{Brownian BSDE for hedger}
\left\{ \begin{array} [c]{ll}
d\widetilde{Y}^{h,x_{1}}_u=\widetilde{Z}^{h,x_{1}}_u\sigma (u, S_{u})\, d\widetilde{W}_u +
G^{h}(u,x_{1}, S_{u}, \widetilde{Y}^{h,x_{1}}_u,\widetilde{Z}^{h,x_{1}}_u)\, du ,\medskip\\
\widetilde{Y}^{h,x_{1}}_T =H(S_{T}),
\end{array} \right.
\ee
where for $x_{1}\ge0$,
\bde
G^{h}(u,x_{1},\ssy,y,z):=z\beta(u,\ssy)-x_{1}\rll_u\Blr_u+\rll_u \Big(y+x_{1}\Blr_u- z\ssy\Big)^+- \rbb_t \Big( y+x_{1}\Blr_u-z\ssy\Big)^-
\ede
and for $x_{1}\leq0$
\bde
G^{h}(u,x_{1},\ssy,y,z):=z\beta(u,\ssy)-x_{1}\rbb_u\Bbr_u+\rll_u \Big(y+x_{1}\Bbr_u- z\ssy\Big)^+- \rbb_u \Big( y+x_{1}\Bbr_u-z\ssy\Big)^-.
\ede
The unique replicating strategy for the hedger equals $\phi = \big(\xi, \psi^{l}, \psi^{b}\big)$ where for every $u\in[t,T]$
$\xi_{u}= \widetilde{Z}^{h,x_{1}}_{u}$ and
\bde
\begin{array} [c]{ll}
\psi^{l}_u = (\Blr_u)^{-1} \Big( \widetilde{Y}^{h,x_{1}}_{u}+x_{1}\Blr_u\I_{\{x_{1}\ge0\}}+x_{1}\Bbr_u\I_{\{x_{1}\leq0\}}- \xi_u S_u \Big)^+, \medskip\\
\psi^{b}_u = - (\Bbr_u)^{-1} \Big( \widetilde{Y}^{h,x_{1}}_{u}+x_{1}\Blr_u\I_{\{x_{1}\ge0\}}+x_{1}\Bbr_u\I_{\{x_{1}\leq0\}}- \xi_u S_u \Big)^-.
\end{array}
\ede

For a fixed $(t,\ssx)\in [0,T) \times \mathcal{O}$, the solution $(\widetilde{Y}^{h,x_{1}},\widetilde{Z}^{h,x_{1}})$ depends on the initial value $\ssx $ of the stock price at time $t$, so that we write $(\widetilde{Y}^{h,x_{1},\ssx},\widetilde{Z}^{h,x_{1},\ssx})$. If we denote $(Y^{h,x_{1},\ssx}_{u},Z^{h,x_{1},\ssx}_{u}):=(\widetilde{Y}^{h,x_{1},\ssx}_{u},\widetilde{Z}^{h,x_{1},\ssx}_{u}\sigma(u,\Sst_{u}))$ and
\bde
\overline{G}^{h}(u,x_{1},\ssy,y,z)=G^{h}(u,x_{1}, \ssy , y,z\sigma^{-1}(u, x)),
\ede
then BSDE (\ref{Brownian BSDE for hedger}) reduces to
\be \label{Brownian BSDE 2 for hedger}
\left\{ \begin{array}
[c]{ll}
dY^{h,x_{1},\ssx}_u = Z^{h,x_{1},\ssx}_u\, d\widetilde{W}_u+\overline{G}^{h}(u,x_{1}, \Sst_{u},Y^{h,x_{1},\ssx}_u,Z^{h,x_{1},\ssx}_u)\, du,\medskip\\
Y^{h,x_{1},\ssx}_T = H( \Sst_{T}).
\end{array} \right.
\ee
Using the non-linear Feynman-Kac formula, under suitable smoothness conditions of the coefficients $\mu,\sigma,\kappa$ and $\beta$,
we deduce that the hedger's pricing function $v(t,\ssx):=Y_{t}^{h,x_{1},\ssx}$ belongs to the class $C^{1,2}([0,T]\times\mathcal{O})$ and solves the following pricing PDE
\bde
\left\{ \begin{array}
[c]{ll}
\frac{\partial v}{\partial t}(t,\ssx)+\mathcal{L}v(t,\ssx)=\overline{G}^{h}\big(t,x_{1},\ssx,v(t,\ssx),\sigma(t,\ssx)\frac{\partial v}{\partial \ssx}(t,\ssx)\big),\quad (t,\ssx)\in[0,T]\times\mathcal{O},\medskip\\
v(T,\ssx)=H(\ssx), \quad \ssx\in\mathcal{O} ,
\end{array} \right.
\ede
where
\bde
\mathcal{L}:=\frac{1}{2}\sigma^{2}(t,\ssx)\frac{\partial^{2} }{\partial \ssx^{2}}+(\beta-\kappa)(t,\ssx)\frac{\partial}{\partial \ssx}.
\ede
Equivalently, the function $v(t,\ssx)$ satisfies
\be \label{Bergman model hedger PDE}
\left\{
\begin{array} [c]{ll}
\frac{\partial v}{\partial t}(t,\ssx)+\frac{1}{2}\sigma^{2}(t,\ssx)\frac{\partial^{2}v }{\partial \ssx^{2}}(t,\ssx)=
\kappa(t,\ssx) \frac{\partial v}{\partial \ssx}(t,\ssx)-x_{1}\rll_t\Blr_t\I_{\{x_{1}\ge0\}}-x_{1}\rbb_t\Bbr_t\I_{\{x_{1}\leq0\}}\medskip\\
\quad \mbox{} +\rll_t \Big(v(t,\ssx)+x_{1}\Blr_t\I_{\{x_{1}\ge0\}}+x_{1}\Bbr_t\I_{\{x_{1}\leq0\}}- \ssx \frac{\partial v}{\partial \ssx}(t,\ssx)\Big)^+\medskip\\
\quad \mbox{} - \rbb_t \Big(v(t,\ssx)+x_{1}\Blr_t\I_{\{x_{1}\ge0\}}+x_{1}\Bbr_t\I_{\{x_{1}\leq0\}}- \ssx \frac{\partial v}{\partial \ssx}(t,\ssx)\Big)^-,\quad (t,\ssx)\in[0,T]\times\mathcal{O},\medskip\\
v(T,\ssx)=H(\ssx), \quad \ssx\in\mathcal{O}.
\end{array}
\right.
\ee

Conversely, if a function $v \in C^{1,2}([0,T]\times\mathcal{O})$ solves PDE (\ref{Bergman model hedger PDE}), then $(v(u,S_{u}),\sigma(u,S_{u})\frac{\partial v}{\partial \ssx}(u,S_{u}))$ solves BSDE (\ref{Brownian BSDE 2 for hedger}) on $u\in[t,T]$ where we write $S = \Sst$. Therefore, $(v(u,S_{u}),\frac{\partial v}{\partial \ssx}(u,S_{u}))$ solves BSDE (\ref{Brownian BSDE for hedger}). Consequently, the unique replicating strategy for the hedger equals $\phi = \big(\xi, \psi^{l}, \psi^{b}\big)$ where for $u\in [t,T]$
\be \label{Bergman model replicating strategy for the hedger}
\begin{array}
[c]{ll}
\xi_{u} =\frac{\partial v}{\partial \ssx}(u,S_{u}),\medskip\\
\psi^{l}_u = (\Blr_u)^{-1} \Big(v(u,S_{u})+x_{1}\Blr_u\I_{\{x_{1}\ge0\}}+x_{1}\Bbr_u\I_{\{x_{1}\leq0\}}- S_u\frac{\partial v}{\partial \ssx}(u,S_{u}) \Big)^+, \medskip\\
\psi^{b}_u = - (\Bbr_u)^{-1} \Big(v(u,S_{u})+x_{1}\Blr_u\I_{\{x_{1}\ge0\}}+x_{1}\Bbr_u\I_{\{x_{1}\leq0\}}-S_u \frac{\partial v}{\partial \ssx}(u,S_{u})\Big)^-.
\end{array}
\ee

Let us now consider the pricing problem for the counterparty with an initial endowment $x_2$. We now have $P^{c}(x_{2},-A,-C)=\widetilde{Y}^{c,x_{2}}$,
where  $(\widetilde{Y}^{c,x_{2}},\widetilde{Z}^{c,x_{2}})$ is the unique solution of following BSDE
\be\label{Brownian BSDE for counterparty}
\left\{
\begin{array}
[c]{ll}
d\widetilde{Y}^{c,x_{2}}_u =\widetilde{Z}^{c,x_{2}}_u\sigma (u, S_{u})\, d\widetilde{W}_u+G^{c}(u,x_{2},S_{u}, \widetilde{Y}^{c,x_{2}}_u,\widetilde{Z}^{c,x_{2}}_u)\, du,\medskip\\
\widetilde{Y}^{c,x_{2}}_T=H(S_{T}),
\end{array}
\right.
\ee
where for $x_{2}\ge0$,
\bde
G^{c}(u,x_{2},\ssy,y,z):=z\beta(u,\ssy)+x_{2}\rll_u\Blr_u-\rll_u \Big(-y+x_{2}\Blr_u+z\ssy\Big)^++\rbb_u \Big(-y+x_{2}\Blr_u+z\ssy\Big)^-
\ede
and for $x_{2}\leq0$
\bde
G^{c}(u,x_{2},\ssy,y,z):=z\beta(u,\ssy)+x_{2}\rbb_u\Bbr_u-\rll_u \Big(-y+x_{2}\Bbr_u+z\ssy\Big)^++\rbb_u \Big(-y+x_{2}\Bbr_u+z\ssy\Big)^-.
\ede
The unique replicating strategy for the counterparty equals $\phi = \big(\xi, \psi^{l}, \psi^{b}\big)$ where, for every $u\in[t,T]$,
$\xi_{u}= -\widetilde{Z}^{c,x_{2}}_{u}$ and
\bde
\begin{array}
[c]{ll}
\psi^{l}_u = (\Blr_u)^{-1} \Big( -\widetilde{Y}^{h,x_{2}}_{u}+x_{2}\Blr_u\I_{\{x_{2}\ge0\}}+x_{2}\Bbr_u\I_{\{x_{2}\leq0\}}- \xi_u S_u \Big)^+, \medskip\\
\psi^{b}_u = - (\Bbr_u)^{-1} \Big( -\widetilde{Y}^{h,x_{2}}_{u}+x_{2}\Blr_u\I_{\{x_{2}\ge0\}}+x_{2}\Bbr_u\I_{\{x_{2}\leq0\}}- \xi_u S_u \Big)^-.
\end{array}
\ede

For a fixed $(t,\ssx)\in [0,T) \times \mathcal{O}$,
we denote $(Y^{c,x_{2},\ssy}_{u},Z^{c,x_{2},\ssy}_{u}):=(\widetilde{Y}^{c,x_{2}}_{u},\widetilde{Z}^{c,x_{2}}_{u}\sigma(u,\Sst_{u}))$ and
\bde
\overline{G}^{c}(u,x_{2},\ssy, y,z)=G^{c}(u,x_{2},\ssy, y,z\sigma^{-1}(u, \ssy)).
\ede
Then BSDE (\ref{Brownian BSDE for hedger}) reduces to
\be\label{Brownian BSDE 2 for counterparty}
\left\{
\begin{array}
[c]{ll}
dY^{c,x_{2},\ssy}_u =Z^{c,x_{2},\ssy}_u\, d\widetilde{W}_u
+\overline{G}^{c}( u,x_{2}, \Sst_{u}, Y^{c,x_{2},\ssy}_u,Z^{c,x_{2},\ssy}_u)\,du,\medskip\\
Y^{c,x_{2},\ssy}_T=H(\Sst_{T}).
\end{array}
\right.
\ee

Under suitable smoothness conditions imposed on the coefficients $\mu$ and $\sigma$, from the Feynman-Kac formula, we deduce that
the function $v(t,\ssx):=Y_{t}^{c,x_{2},\ssx}$ belongs to $C^{1,2}([0,T]\times\mathcal{O})$
and solves the following PDE
\be\label{Bergman model counterparty PDE 1}
\left\{
\begin{array}
[c]{ll}
\frac{\partial v}{\partial t}(t,\ssx)+\mathcal{L}v(t,\ssx)=\overline{G}^{c}\big(t,x_{2},\ssx,v(t,\ssx),\sigma(t,\ssx)\frac{\partial v}{\partial \ssx}(t,\ssx)\big), \quad (t,\ssx)\in[0,T]\times\mathcal{O},\medskip\\
v(T,\ssx)=H(\ssx), \quad \ssx\in\mathcal{O}.
\end{array}
\right.
\ee
More explicitly,
\be\label{Bergman model counterparty PDE}
\left\{
\begin{array}
[c]{ll}
\frac{\partial v}{\partial t}(t,\ssx)+\frac{1}{2}\sigma^{2}(t,\ssx)\frac{\partial^{2}v }{\partial \ssx^{2}}(t,\ssx)=
\kappa(t,\ssx) \frac{\partial v}{\partial \ssx}(t,\ssx)+x_{2}\rll_t\Blr_t\I_{\{x_{2}\ge0\}}+x_{2}\rbb_t\Bbr_t\I_{\{x_{2}\leq0\}}\medskip\\
\quad \mbox{} -\rll_t \Big(-v(t,\ssx)+x_{2}\Blr_t\I_{\{x_{2}\ge0\}}+x_{2}\Bbr_t\I_{\{x_{2}\leq0\}}+ \ssx \frac{\partial v}{\partial \ssx}(t,\ssx)\Big)^+\medskip\\
\quad \mbox{} + \rbb_t \Big(-v(t,\ssx)+x_{2}\Blr_t\I_{\{x_{2}\ge0\}}+x_{2}\Bbr_t\I_{\{x_{2}\leq0\}}+ \ssx \frac{\partial v}{\partial \ssx}(t,\ssx)\Big)^-,\quad (t,\ssx)\in[0,T]\times\mathcal{O},\medskip\\
v(T,\ssx)=H(\ssx), \quad \ssx\in\mathcal{O}.
\end{array}
\right.
\ee

Conversely, if a function $v\in C^{1,2}([0,T]\times\mathcal{O})$ solves PDE (\ref{Bergman model counterparty PDE}), then  $(v(u,S_{u}),\sigma(u,S_{u})\frac{\partial v}{\partial \ssx}(u,S_{u}))$ solves BSDE (\ref{Brownian BSDE 2 for counterparty}) on $u\in[t,T]$ where $S = \Sst$. Hence $(v(u,S_{u}),\frac{\partial v}{\partial \ssx}(u,S_{u}))$ solves BSDE (\ref{Brownian BSDE for counterparty})  and the unique replicating strategy for the counterparty equals $\phi = \big(\xi, \psi^{l}, \psi^{b}\big)$ where, for every $u\in [t,T]$,
\be \label{Bergman model replicating strategy for the counterparty}
\begin{array}
[c]{ll}
\xi_{u}=-\frac{\partial v}{\partial \ssx}(u,S_{u}),\medskip\\
\psi^{l}_u = (\Blr_u)^{-1} \Big(-v(u,S_{u})+x_{2}\Blr_u\I_{\{x_{2}\ge0\}}+x_{2}\Bbr_u\I_{\{x_{2}\leq0\}}+S_u \frac{\partial v}{\partial \ssx}(u,S_{u}) \Big)^+, \medskip\\
\psi^{b}_u = - (\Bbr_t)^{-1} \Big(-v(u,S_{u})+x_{2}\Blr_u\I_{\{x_{2}\ge0\}}+x_{2}\Bbr_u\I_{\{x_{2}\leq0\}}+ S_u \frac{\partial v}{\partial \ssx}(u,S_{u}) \Big)^-.
\end{array}
\ee

The following proposition summarizes the above considerations. When $\kappa=0$ (that is, the stock pays no dividends) and $x_{1}=x_{2}=0$, then PDE (\ref{Bergman model hedger PDE}) reduces to PDE (5) in Bergman \cite{B-1995}. Therefore, Proposition \ref{Bergman model pricing using PDE} can be seen as a generalization of Proposition 2 in \cite{B-1995}.

\bp \label{Bergman model pricing using PDE}
If $v(t,\ssx)\in  C^{1,2}([0,T]\times\mathcal{O})$ is the solution of quasi-linear PDE (\ref{Bergman model hedger PDE}),
then the hedger's ex-dividend price of the European claim $H(S_T)$ equals $v(t,S_{t})$ and the unique
replicating strategy $\phi = \big(\xi, \psi^{l}, \psi^{b}\big)$ for the hedger is given by (\ref{Bergman model replicating strategy for the hedger}). Similarly, if $v(t,\ssx)\in  C^{1,2}([0,T]\times\mathcal{O})$ is the solution of quasi-linear PDE (\ref{Bergman model counterparty PDE}), then the counterparty's ex-dividend price of the European claim $H(S_T)$ equals $v(t,S_{t})$ and the unique replicating strategy $\phi = \big(\xi, \psi^{l}, \psi^{b}\big)$ for the counterparty is  given by (\ref{Bergman model replicating strategy for the counterparty}).
\ep

\vskip 5 pt


\noindent {\bf Acknowledgement.}
The research of Tianyang Nie and Marek Rutkowski was supported under Australian Research
Council's Discovery Projects funding scheme (DP120100895).




\begin{thebibliography}{20}
{\parskip = 3 pt

\bibitem{B-1995}
Bergman, Y. Z.:
Option pricing with differential interest rates.
{\it Review of Financial Studies} 8 (1995), 475--500.

\bibitem{BR-2014}
Bielecki, T. R., Rutkowski, M.:
Valuation and hedging of contracts with funding costs and collateralization.
Working paper, 2014.

\bibitem{BCPP11}
Brigo, D., Capponi, A., Pallavicini, A., Papatheodorou, V.:
Collateral margining in arbitrage-free counterparty valuation adjustment including re-hypothecation and netting.
Working paper, 2011.

\bibitem{BK09}
Burgard, C., Kjaer, M.:
PDE representations of options with bilateral counterparty risk and funding costs.
Working paper, November, 2009.

\bibitem{BK11}
Burgard, C., Kjaer, M.:
Partial differential equations representations of derivatives with counterparty risk and funding costs.
{\it Journal of Credit Risk} 7 (2011), 1--19.


\bibitem{SC12a}
Cr\'{e}pey, S.: Bilateral counterparty risk under funding constraints -- Part I: Pricing.
\textit{Mathematical Finance} (published online on 12 December 2012).

\bibitem{SC12b}
Cr\'{e}pey, S.: Bilateral counterparty risk under funding constraints -- Part II: CVA.
\textit{Mathematical Finance} (published online on 12 December 2012).



\bibitem{EPQ-1997}
El Karoui, N., Peng, S., Quenez, M. C.: Backward stochastic differential equations in finance.
{\it Mathematical Finance} 7 (1997), 1--71.


%

\bibitem{M-2013} Mercurio, F.:
Bergman, Piterbarg and beyond: Pricing derivatives under collateralization
and differential rates. Working paper, 2013.

%

\bibitem{NR2}
Nie, T., Rutkowski, M.: Fair and profitable bilateral prices under funding costs and collateralization.
Working paper, University of Sydney, 2014.

\bibitem{NR3}
Nie, T., Rutkowski, M.: BSDEs driven by a multi-dimensional martingale and their applications to market models with funding costs. Working paper, University of Sydney, 2014.

\bibitem{PPB12}
Pallavicini, A., Perini, D., Brigo, D.:
Funding, collateral and hedging: uncovering the mechanics and the subtleties of funding valuation adjustments.
Working paper, 2012.





\bibitem{P10}
Piterbarg, V.:
Funding beyond discounting: collateral agreements and derivatives pricing.
\textit{Risk}, February (2010), 97--102.

}
\end{thebibliography}
\end{document}